\documentclass[10pt,a4paper,oneside]{article}

\usepackage{amssymb}
\usepackage{amsmath}
\usepackage{amsthm}

\makeatletter
\let\@afterindentfalse\@afterindenttrue
\@afterindenttrue
\makeatother

\newtheorem{theorem}{Theorem}[section]
\newtheorem{corollary}{Corollary}[section]

\newcommand{\MM}{\mathcal{M}^{+}_{2}}

\newcommand{\MN}{\mathcal{M}^{+}_{n}}
\newcommand{\MNm}{\mathcal{M}^{+}_{m}}
\newcommand{\TMN}{\mathcal{M}_{n}}
\newcommand{\Tr}{\mathop{\textrm{Tr}}\nolimits}
\newcommand{\Ric}{\mathop{\textrm{Ric}}\nolimits}
\newcommand{\Scal}{\mathop{\textrm{Scal}}\nolimits}
\newcommand{\LIN}{\mathop{\textrm{LIN}}\nolimits}
\newcommand{\la}{\lambda}

\begin{document}

\title{Monotone Riemannian metrics on density matrices \\
  with non-monotone scalar curvature
   \thanks{keywords: state space,
                   monotone statistical metric, scalar curvature;
           PACS numbers: 06.67.-a, 02.40.Ky, 02.50.-r}}
\author{Attila Andai\thanks{andaia@math.bme.hu}\\
  Department for Mathematical Analysis, \\
  Budapest University of Technology and Economics,\\
  H-1521 Budapest XI. Sztoczek u. 2, Hungary}
\date{May 18, 2003}

\maketitle

\begin{abstract}
  The theory of monotone Riemannian metrics on the state space of a
    quantum system was established by D\'enes Petz in 1996.
  In a recent paper he argued that the scalar curvature of a statistically
    relevant -- monotone -- metric can be interpreted as an average
    statistical uncertainty.
  The present paper contributes to this subject.
  It is reasonable to expect that states which are more mixed are less
   distinguishable than those which are less mixed.
  The manifestation of this behavior could be that for such a metric the
    scalar curvature has a maximum at the maximally mixed state.
  We show that not every monotone metric fulfils this expectation,
    some of them behave in a very different way.
  A mathematical condition is given for monotone Riemannian metrics
    to have a local minimum at the maximally mixed state and examples are
    given for such metrics.
\end{abstract}
\bigskip

\section{INTRODUCTION}

The quantum mechanical Hilbert space formalism gives a mathematical
  description of particles with spin of $\frac{n-1}{2}$.
Concentrating on the spin part of non relativistic particles one
  can build a proper mathematical model in an $n$ dimensional complex
  Hilbert space.
This is the simplest physical realization of an $n$-level quantum system.
The states of an $n$-level system are identified with the set
  of positive semidefinite self-adjoint $n\times n$ matrices of trace 1.
The states form a closed convex set in the space of matrices and its interior,
  the set of all strictly positive self-adjoint matrices of trace 1 becomes
  naturally a differentiable manifold.

The idea in mathematical statistics that a statistical or informational distance
  between probability measures gives rise to a Riemannian metric is due to Rao$^{1}$
  and was developed by Amari$^{2}$ and Streater$^{3}$ among others.
Let us see how can one measure the statistical distance between the simplest
  probability distributions in classical case Ref. 4.
Assume, that we have two probability distributions $(p_{1},1-p_{1})$ and $(p_{2},1-p_{2})$.
Let us now suppose that an experimenter, in making his determination of the value $p_{1}$,
  has only $n$ trials.
Because of the unavoidable statistical fluctuations associated with a finite sample,
  the experimenter cannot know $p_{1}$ exactly.
After these measurements the experimenter's uncertainty is the size of a typical
  fluctuation
  $$\Delta p_{1}=\sqrt{\frac{p_{1}(1-p_{1})}{n}}.$$
We can say that the distributions $(p_{1},1-p_{1})$ and $(p_{2},1-p_{2})$ are distinguishable
  in $n$ trials if the regions of uncertainty do not overlap,
  that is, if
  $$\vert p_{1}-p_{2}\vert\geq \Delta p_{1}+\Delta p_{2}.$$
Let $k(n,p_{1},p_{2})$ denote the number of the probability distributions of the form $(p_{i},1-p_{i})$
  between $p_{1}$ and $p_{2}$ (that is $p_{1}<p_{i}<p_{2}$) each of which is distinguishable in $n$
  trials from its neighbors.
The statistical distance between the given probability distributions is
  $$d(p_{1},p_{2})=\lim_{n\to\infty}\frac{k(n,p_{1},p_{2})}{\sqrt{n}}.$$
From the previous equations we find that
  $$d(p_{1},p_{2})=\int_{p_{1}}^{p_{2}}\frac{1}{2\sqrt{p(1-p)}}\ dp
  =\arccos\bigl(\sqrt{p_{1} p_{2}}+\sqrt{(1-p_{1})(1-p_{2})} \bigr).$$
This distance function was introduced by Fisher in 1922.

The Fisher informational metric is unique in some sense (i.e., it is the only Markovian
  monotone distance) in the classical case Ref. 5.
A family of Riemannian metrics are called monotone if they are decreasing under
  stochastic mappings (the exact definition is given below).
These metrics play the role of Fisher metric in the quantum case.
Monotone Riemannian metrics are important for information-theoretical and
  statistical considerations on the state space.
The study of monotone metrics for parametric statistical manifolds was
  initiated by Chentsov and Morozova$^{6}$.
Petz's classification theorem$^{7}$ establishes a correspondence between
  monotone metrics and operator monotone
  $f:\lbrack 0,\infty\lbrack\to\mathbb{R}$ functions, such that $f(x)=xf(x^{-1})$ hold for all
  positive $x$.
In the simplest quantum case, dealing with $2\times 2$ matrices we can use the
  Stokes parametrization, that is every state $D$ can be uniquely written in the
  $$D=\frac{1}{2}\bigl(I+x_{1}\sigma_{1}+x_{2}\sigma_{2}+x_{2}\sigma_{2}\bigr)$$
  form, where $(\sigma_{i})_{i=1,2,3}$ are the Pauli matrices and
  $(x_{1},x_{2},x_{3})\in\mathbb{R}^3$ with $x_{1}^{2}+x_{2}^{2}+x_{3}^{2}\leq 1$.
The interior of the set of states can be identified with the open
unit ball
  in $\mathbb{R}^3$ by this parametrization.
In this case a monotone Riemannian metric on this manifold can be written in the
  $$ds^{2}=\frac{1}{1-r^{2}}dr^{2}
    +\frac{r^{2}}{(1+r)f\left(\frac{1-r}{1+r}\right)} d\vartheta^{2}
    +\frac{r^{2}\sin^{2}\vartheta}{(1+r)f\left(\frac{1-r}{1+r}\right)} d\varphi^{2}$$
  form in polar coordinates.

There is strong a connection between the scalar curvature of these
  manifolds at a given state and statistical distinguishability
  and uncertainty of the state.
If $D_{0}$ is a $n\ \times\ n$  density matrix
  we call geodesic ball the set
$$B_{r}(D_{0})=\lbrace D\ n\ \times\ n\ \mbox{density matrix} :\ d(D_{0},D)< r\rbrace.$$
  The volume of this ball is given by
$$V(B_{r}(D_{0}))=\frac{\sqrt{\pi^{n^{2}-1}}r^{n^{2}-1}}{\Gamma\left(\frac{n^{2}+1}{2}\right)}\cdot
  \left(1-\frac{r(D_{0})}{6(n^{2}+1)}r^{2}+O(r^{4}) \right) $$
  where $r(D_{0})$ is the scalar curvature at the point $D_{0}$.
According to Ref. 8, the quantity $V(B_{r}(D_{0}))$ measures the statistical uncertainty
  and the scalar curvature measures the average statistical uncertainty.
A general explicit formula for the scalar curvature was given by Dittmann$^{10}$
  (particular cases have been discussed in Refs. 11 and 12).
There are many Riemannian metrics on the state space which are
  statistically relevant in different ways.
The state $A$ is more mixed than the state $B$ if for their
  decreasingly ordered set of eigenvalues $(a_{1},\dots,a_{n})$ and
$(b_{1},\dots,b_{n})$
  the inequality
$$\sum_{l=1}^{k}a_{l}\leq\sum_{l=1}^{k}b_{l}$$
  holds for every $1\leq k\leq n$.

It is reasonable to expect that the most mixed
  states are less distinguishable than the less mixed states; for details see
  Refs. 13 and 14.
It means mathematically that in this case the scalar curvature
  of a Riemann structure should have the following monotonicity property:
  if $D_{1}$ is more mixed than $D_{2}$ then
  $r(D_{2})$ should be less then $r(D_{1})$, where $r(D)$ denotes the
  scalar curvature of the manifold at the state $D$.
It has been shown that the Bures metric does not have this monotonicity property
  for the scalar curvature and moreover it has a global minimum at the most
  mixed state.
Actually in Ref. 9 it was proved that for the Bures metric for every $n\times n$
  density matrix $D$ the inequality
$$r(D)\geq\frac{(5n^{2}-4)(n^{2}-1)}{4}$$
  holds.
If $n>3$, equality holds iff $D=\frac{1}{n}I$. If $n=2$, then $r(D)=24$.
This implies that the Bures metric is (trivially)  monotone for $n=2$.
Indeed all metrics studied thus far in two level quantum systems
 have monotone scalar curvature.
This means that these metrics are compatible with one's statistical view.
Do all monotone metrics have this property in two level quantum
  system?
The answer is no; in this paper we show a family of monotone metrics
  with non monotone scalar curvature and we give a condition for
  monotone metric to have a local minimum at the maximally mixed state.

\section{SCALAR CURVATURE ON THE TWO LEVEL QUANTUM SYSTEMS}

\subsection{The setup}

Let $\MN$ be the space of all complex self-adjoint positive
  definite $n\times n$ matrices of trace 1 and
  let $\TMN$ be the real vector space of all self-adjoint traceless
  $n\times n$ matrices.
The space $\MN$ can be endowed with a differentiable structure
Ref. 15.

The tangent space $T_{D}$ at $D\in\MN$ can be identified with $\TMN$.
A map
$$K:\MN\times\TMN\times\TMN\to\mathbb{C}\qquad
    (D,X,Y)\mapsto K_{D}(X,Y)$$
will be called a Riemannian metric if the following condition hold:
For all $D\in\MN$ the map
$$K_{D}:\TMN\times\TMN\to\mathbb{C}\qquad (X,Y)\mapsto K_{D}(X,Y)$$
is a scalar product and for all $X\in\TMN$ the map
$$K_{\cdot}(X,X):\MN\to\mathbb{C} \qquad D\mapsto K_{D}(X,X)$$
is smooth.

We now use differential geometrical notation to define the scalar
curvature of the $(\MN,K)$ Riemannian manifold. In this case the
Riemannian metric is a
 $$K:\MN\to\LIN(\TMN\times\TMN,\mathbb{R})\qquad
  D\mapsto \bigl( (X,Y)\mapsto K_{D}(X,Y)\bigr)$$
 map, where $\LIN(U,V)$ denotes the set of linear maps from the vector space $U$ to the
 vector space $V$.
The derivative of the metric $K$ is a map
  $$dK:\MN\to \LIN \bigl(\TMN,\LIN(\TMN\times\TMN,\mathbb{R})\bigr)\qquad
  D\mapsto \Bigl( X\mapsto\bigl( (Y,Z)\mapsto dK_{D}(X)(Y,Z) \bigr) \Bigr).$$
At a given $D\in\MN$ point for given $X,Y\in\TMN$ tangent vectors the map
$$\tau_{D,X,Y}:\TMN\to\mathbb{R}\qquad Z\mapsto\frac{1}{2}\bigl(
    dK_{D}(Y)(X,Z)+dK_{D}(X)(Z,Y)-dK_{D}(Z)(X,Y)  \bigr)$$
is a linear functional. It means that there exists a unique
$V_{D,X,Y}\in\TMN$ tangent vector such that for all $Z\in\TMN$ vector
$$K_{D}(V_{D,X,Y},Z)=\tau_{D,X,Y}(Z)$$
holds. One can define the map
$$\Gamma:\MN\to\LIN(\TMN\times\TMN,\TMN)\qquad
  D\mapsto \bigl( (X,Y)\mapsto V_{D,X,Y}\bigr)$$
which is called covariant differentiation. Its derivative is a map
$$d\Gamma:\MN\to\LIN\bigl(\TMN,\LIN(\TMN\times\TMN,\TMN) \bigr)\qquad
  D\mapsto \Bigl( X\mapsto\bigl( (Y,Z)\mapsto d\Gamma_{D}(X)(Y,Z) \bigr) \Bigr).$$
The Riemann curvature tensor defined to be
$$R:\MN\to\LIN(\TMN\times\TMN\times\TMN,\TMN)\qquad (D,X,Y,Z)\mapsto R_{D}(X,Y,Z),$$
where
$$R_{D}(X,Y,Z)=d\Gamma_{D}(X)(Y,Z)-d\Gamma_{D}(Y)(X,Z)
  +\Gamma_{D}\bigl(X,\Gamma_{D}(Y,Z) \bigr)-\Gamma_{D}\bigl(Y,\Gamma_{D}(X,Z) \bigr).$$
The map
$$\alpha:\MN\times\TMN\times\TMN\to\LIN(\TMN,\TMN)\qquad
(D,X,Y)\mapsto \alpha_{D,X,Y}=\bigl(Z\mapsto R_{D}(Z,X,Y) \bigr),$$
is needed to define the Ricci tensor
$$\Ric:\MN\to\LIN(\TMN\times\TMN,\mathbb{R})\qquad
  D\mapsto\bigl( (X,Y)\mapsto \Ric_{D}(X,Y) \bigr),$$
where
$$\Ric_{D}(X,Y)=\Tr\alpha_{D,X,Y}.$$
At a given $D\in\MN$ point for given $X\in\TMN$ tangent vector the map
$$\beta_{D,X}:\TMN\to\mathbb{R}\qquad Y\mapsto\Ric_{D}(X,Y)$$
is a linear functional. It means that there exists a unique
$U_{D,X}\in\TMN$ tangent vector such that for all $Y\in\TMN$ vector
$$K_{D}(U_{D,X},Y)=\beta_{D,X}(Y)$$
holds. From the map
$$\rho:\MN\to\LIN(\TMN,\TMN)\qquad
  D\mapsto \rho_{D}=\bigl( (X)\mapsto U_{D,X}\bigr)$$
we get the scalar curvature of the manifold
$$\Scal:\MN\to\mathbb{R} \qquad D\mapsto\Tr\rho_{D}.$$
For further differential geometry details see, for example, Ref. 16.

Let $M_{n}(\mathbb{C})$ denote the set of complex $n\times n$ matrices
and $M_{k}(M_{n}) $ denote the set of $k\times k$ matrices with entries
$M_{n}(\mathbb{C})$. If $T:M_{n}(\mathbb{C})\to M_{m}(\mathbb{C}) $ is a linear map, it
induces a linear map $T^{(k)}:M_{k}(M_{n})\to M_{k}(M_{m})$ by
$$T^{(k)}(\lbrack A_{ij}\rbrack)=\lbrack T(A_{ij})\rbrack.$$
The map $T$ is called positive if it takes positive operators to
positive operators. Say that $T$ is $k$-positive if $T^{(k)}$ is
positive and $T$ is completely positive if it is $k$-positive for all $k\geq 1$.

A linear mapping $T:M_{n}(\mathbb{C})\to M_{m}(\mathbb{C})$ is defined
to be stochastic if $T$ is completely positive and trace preserving.
For more information on completely positive and stochastic maps see Ref. 13 and 14.

Let $(K^{m})_{m\in\mathbb{N}}$ be a family of metrics, such that
$K^{m}$ is a Riemannian metric on $\MNm$ for all $m$.
This family of metrics defined to be monotone if
$$K_{T(D)}^{m}(T(X),T(X))\leq K_{D}^{n}(X,X)$$
for every stochastic mapping $T:M_{n}(\mathbb{C})\to M_{m}(\mathbb{C})$,
for every $D\in\MN$ and for all $X\in\TMN$ and for all
$m,n\in\mathbb{N}$.

\begin{theorem}\label{th:1}
{\it Petz classification theorem$^{7}$:}
There exists a bijective correspondence between the monotone
family of metrics $(K^{n})_{n\in\mathbb{N}}$ and operator monotone
$f:\mathbb{R}^{+}\to\mathbb{R}$ functions such that
$f(x)=xf(x^{-1})$ hold for all positive $x$.
The metric is given by
\begin{equation}K_{D}^{n}(X,Y)=\Tr\Bigl(X\bigl(
     R_{n,D}^{\frac{1}{2}} f(L_{n,D}R_{n,D}^{-1}) R_{n,D}^{\frac{1}{2}}
    \bigr)^{-1}(Y)\Bigr)\label{eq:Petz repr}
\end{equation}
for all $n\in\mathbb{N}$ where $L_{n,D}(X)=DX$, $R_{n,D}(X)=XD$ for
all $D,X\in M_{n}(\mathbb{C})$.
\end{theorem}

A Riemannian metric $K$ is said to be monotone if there is a
monotone family of metrics $(K^{m})_{m\in\mathbb{N}}$ such that
$K=K^{n}$ for an $n$. We use the normalization condition
for the function $f$ in the previous theorem $f(1)=1$.
Here are some examples of operator monotone functions which
generate monotone metrics from Refs. 17 and 18:
$$\frac{1+x}{2},\ \frac{2x}{1+x},\ \frac{x-1}{\log x},
  \ \frac{2(x-1)^{2}}{(1+x)(\log x)^{2}},
  \ \frac{2(x-1)\sqrt{x}}{(1+x)\log x},
  \ \frac{2x^{\alpha+1/2}}{1+x^{2\alpha}},
  \ \frac{\beta(1-\beta)(x-1)^{2}}{(x^{\beta}-1)(x^{1-\beta}-1)},$$
where $0\leq\alpha\leq 1/2$ and $0<\vert\beta\vert< 1$.

\subsection{Curvature and eigenvalues on $\MM$}

There is an explicit formula for scalar curvature in a given
$D\in\MM$ state using a monotone metric coming from a suitable $f$
function defined by Eq. (1). To use that result to build up a more
explicit formula to our $\MM$ manifold, first introduce the
Morozova-Chentsov function related to the monotone function $f$
defined by
\begin{equation}
 c(x,y):=\frac{1}{yf(x/y)}.\label{eq:c def}
\end{equation}
Let us denote by $\partial_{1}c(x,y)$ the partial derivative of
$c(x,y)$ with respect to its first variable. Define four new
functions (as in Ref. 10)
\begin{eqnarray}
h_{1}(x,y,z):=\dfrac{c(x,y)-zc(x,z)c(y,z)}{(x-z)(y-z)c(x,z)c(y,z)}
\ & h_{3}(x,y,z):=\dfrac{z}{x-y}
      \left(\partial_{1}(\log c)(z,x)-\partial_{1}(\log c)(z,y) \right) \nonumber\\[1em]
h_{2}(x,y,z):=\dfrac{(c(x,z)-c(y,z))^{2}}{(x-y)^{2}c(x,y)c(x,z)c(y,z)}
\ & h_{4}(x,y,z):=z\partial_{1}(\log c)(z,x)\partial_{1}(\log
c)(z,y). \label{eq:hi def}
\end{eqnarray}

The terms like $h_{i}(x,x,z)$ can be computed as a
$$\lim_{y\to x} h_{i}(x,y,z)$$
limit.
We will need a linear combination of these functions
\begin{equation}
h(x,y,z)=h_{1}(x,y,z)-\frac{1}{2}h_{2}(x,y,z)+2h_{3}(x,y,z)-h_{4}(x,y,z).
 \label{eq:h def }
\end{equation}

\begin{theorem} (see Ref. 10)
Let $\sigma(D)$ be the spectrum of the state $D\in\MN$.
Then for the scalar curvature one has the expression
$$r(D)=\sum_{x,y,z\in\sigma(D)}h(x,y,z)-\sum_{x\in\sigma(D)}h(x,x,x)
  +\frac{1}{4}(n^{2}-1)(n^{2}-2).$$
\end{theorem}

\begin{corollary}
The scalar curvature at the state $D\in\MM$ with eigenvalues
$\la_{1},\la_{2}$ is given by
$$r(D)=
h(\la_{1},\la_{1},\la_{2})+h(\la_{1},\la_{2},\la_{1})+h(\la_{2},\la_{1},\la_{1})+
h(\la_{2},\la_{2},\la_{1})+h(\la_{2},\la_{1},\la_{2})+h(\la_{1},\la_{2},\la_{2})+\frac{3}{2}.$$
\end{corollary}

\begin{theorem} Let $D\in\MM$ and $a=2\la_1-1$ where $\la_{1}$ is an
eigenvalue of $D$ and assume that the monotone metric of $\MM$ comes from
a function $f$.
Then the scalar curvature at $D$ is
\begin{eqnarray}
r(a)=& \dfrac
 {14(a-1)\left\lbrack f'\left(\frac{1-a}{1+a}\right)\right\rbrack^{2}}
 {(1+a)^{3}\left\lbrack f\left(\frac{1-a}{1+a}\right)\right\rbrack^{2}}
       +\dfrac{2(a^{2}+7a-6)f'\left(\frac{1-a}{1+a}\right)}
         {(1+a)^{2}af\left(\frac{1-a}{1+a}\right)}
       +\dfrac{8(1-a)f''\left(\frac{1-a}{1+a}\right)}
         {(1+a)^{3}f\left(\frac{1-a}{1+a}\right)} \label{eq:r(a)} \\
     & +\dfrac{2(1+a)f\left(\frac{1-a}{1+a}\right)}{a^{2}}
       +\dfrac{3a^{3}+5a^{2}+8a-4}{2(1+a)a^{2}}. \nonumber
\end{eqnarray}
\end{theorem}

\begin{proof}[Proof 1]

Through the computation we will use the identities
$f'(1)=\frac{1}{2}$ and $2f^{(3)}(1)+3f^{(2)}(1)=0$ which come
from the equations $f(x)=xf(1/x)$ and $f(1)=1$. It is easy to recognize
that $h_{i}(y,x,x)=h_{i}(x,y,x)$ for $i=1,2,3,4$. First let us
note the following identities
\begin{align}
&c(x,x)=\frac{1}{x},
  &\quad &c(x,y)=c(y,x),
  &\quad & c(x,y)=tc(tx,ty),\quad \forall t\in\mathbb{R}^{+},  \label{eq:c properties}\\
&\partial_{1}c(x,x)=-\frac{1}{2x^{2}},
  &\quad &\partial_{1}^{k}\partial_{2}^{l}c(x,y)=\partial_{1}^{l}\partial_{2}^{k}c(y,x),
  &\quad &c(x,y)=-x\partial_{1}c(x,y)-y\partial_{2}c(x,y) \nonumber
\end{align}
which will be used through the computation.

The $h_{i}(x,x,y)$ and $h_{i}(x,y,x)$ like limit functions can be
computed. For example:
$$\begin{array}{rl}
h_{1}(x,x,y)&\displaystyle=\lim_{q\to x}h_{1}(x,q,y)
  =\lim_{q\to x}\frac{c(x,q)-yc(x,y)c(q,y)}{(x-y)(q-y)c(x,y)c(q,y)}
  =\frac{c(x,x)-y\left\lbrack c(x,y)\right\rbrack^{2}}
    {(x-y)^{2}\left\lbrack c(x,y)\right\rbrack^{2}},\\[1em]
h_{4}(x,y,x)&\displaystyle=\lim_{q\to x}h_{4}(x,y,q)
  =\lim_{q\to x}q\frac{\partial_{1}c(q,x)}{c(q,x)}\frac{\partial_{1}c(q,y)}{c(q,y)}
  =x\frac{\partial_{1}c(x,x)}{c(x,x)}\frac{\partial_{1}c(x,y)}{c(x,y)}.
\end{array}$$
After taking into account the identities (\ref{eq:c properties})
these limit functions can be simplified:
\begin{equation}
h_{1}(x,x,y)=
  \dfrac{1-xy\left\lbrack c(x,y)\right\rbrack^{2}}
        {x(x-y)^{2}\left\lbrack c(x,y)\right\rbrack^{2}} \qquad
h_{1}(x,y,x)=-\frac{1}{2}\dfrac{c(x,y)+2x\partial_{1}c(x,y)}{(x-y)c(x,y)}\label{eq:hi}
\end{equation}
\begin{equation}
h_{2}(x,x,y)=x\left(\dfrac{\partial_{1}c(x,y)}{c(x,y)}\right)^{2}\qquad
h_{2}(x,y,x)=\dfrac{1}{x}\left(\dfrac{1-xc(x,y)}{(x-y)c(x,y)}\right)^{2}\nonumber
\end{equation}
$$\begin{array}{rl}
h_{3}(x,x,y)&=-\dfrac
 {y^{2}c(x,y)\left\lbrack\partial_{1}c(y,x)\right\rbrack^{2}
  +2yc(x,y)\partial_{1}c(y,x)
  +xy\partial_{1}c(x,y)\partial_{1}c(y,x)}
 {x\left\lbrack c(x,y)\right\rbrack^{2}}\\
h_{3}(x,y,x)&=-\dfrac{c(x,y)+2x\partial_{1}c(x,y)}{2(x-y)c(x-y)}\\
\end{array}$$
$$\begin{array}{rlrl}
h_{4}(x,x,y)&=y\left(\dfrac{\partial_{1}c(y,x)}{c(x,y)}\right)^{2}
\qquad & h_{4}(x,y,x)&=-\dfrac{1}{2}\dfrac{\partial_{1}c(x,y)}{c(x,y)}.
\end{array}$$

Introducing the suitable sum-functions for $h_{i}$ ($i=1,2,3,4$)
\begin{equation}
sh_{i}(x,y):=h_{i}(x,x,y)+2h_{i}(x,y,x)+h_{i}(y,y,x)+2h_{i}(y,x,y)
 \label{eq:shi def}
\end{equation}
 we get that
\begin{align}
 &sh_{1}(x,y)=\dfrac{(x+y)(1-xy\left\lbrack c(x,y)\right\rbrack^{2})}
                    {xy(x-y)^{2}\left\lbrack c(x,y)\right\rbrack^{2}}
   -\dfrac{4x\partial_{1}c(x,y)+2c(x,y)}{(x-y)c(x,y)}\label{eq:shi(c)}\\[1em]
 &sh_{2}(x,y)=\dfrac{ x\left\lbrack\partial_{1}c(x,y)\right\rbrack^{2}
                     +y\left\lbrack\partial_{1}c(y,x)\right\rbrack^{2}}
                    {\left\lbrack c(x,y)\right\rbrack^{2}}
   +2\dfrac{(x+y)+xy(x+y)\left\lbrack c(x,y)\right\rbrack^{2}
           -4xyc(x,y)}{xy(x-y)^{2}\left\lbrack c(x,y)\right\rbrack^{2}}\nonumber\\[1em]
 &sh_{3}(x,y)=\dfrac{(x+y)\bigl(c(x,y)\left\lbrack\partial_{1,2}c(x,y)\right\rbrack^{2}-
                    \partial_{1}c(x,y)\partial_{1}c(y,x)\bigr)}
                   {\left\lbrack c(x,y)\right\rbrack^{2}}
   -\dfrac{4x\partial_{1}c(x,y)+2c(x,y)}{(x-y)c(x,y)}\nonumber\\[1em]
 &sh_{4}(x,y)=\dfrac{ x\left\lbrack\partial_{1}c(x,y)\right\rbrack^{2}
                     +y\left\lbrack\partial_{1}c(y,x)\right\rbrack^{2}}
                    {\left\lbrack c(x,y)\right\rbrack^{2}}
   -\dfrac{\partial_{1}c(x,y)+\partial_{1}c(y,x)}{c(x,y)}\nonumber
\end{align}

These sum-functions can be expressed by the operator monotone
function $f(x)$:
\begin{align}
 sh_{1}(x,y)&=\dfrac{1}{(x-y)^{2}}\left(\dfrac{y(x+y)}{x}\left\lbrack f(x/y)\right\rbrack^{2}+y-3x
             +\dfrac{4x(x-y)}{y}\dfrac{f'(x/y)}{f(x/y)} \right)\label{eq:shi(f)}\\[1em]
sh_{2}(x,y)&=\dfrac{x}{y^{2}}\left(\dfrac{f'(x/y)}{f(x/y)}\right)^{2}
             +\dfrac{y^{3}}{x^{4}}
              \left(\dfrac{f(x/y)f'(y/x)}{\left\lbrack f(y/x)\right\rbrack^{2}}\right)^{2}
             +\dfrac{2y(x+y)}{x(x-y)^{2}}\left\lbrack f(x/y)\right\rbrack^{2}\nonumber\\[1em]
           & -\dfrac{8y}{(x-y)^{2}}f(x/y)+\dfrac{2(x+y)}{(x-y)^{2}}\nonumber\\[1em]
sh_{3}(x,y)&=\dfrac{-2x(x+y)}{y^{3}}\left(\dfrac{f'(x/y)}{f(x/y)}\right)^{2}
             -\dfrac{(x+y)}{x^{2}}\dfrac{f'(x/y)f'(y/x)}{\left\lbrack f(x/y)\right\rbrack^{2}}
             +\dfrac{2(x^{2}+2xy-y^{2})}{y^{2}(x-y)}\dfrac{f'(x/y)}{f(x/y)}\nonumber\\[1em]
           & +\dfrac{x(x+y)}{y^{3}}\dfrac{f''(x/y)}{f(x/y)}
             -\dfrac{2}{x-y}\nonumber \\[1em]
sh_{4}(x,y)&=\dfrac{x}{y^{2}}\left(\dfrac{f'(x/y)}{f(x/y)}\right)^{2}
             +\dfrac{y^{3}}{x^{4}}\left(\dfrac{f(x/y)f'(y/x)}
                    {\left\lbrack f(y/x)\right\rbrack^{2}}\right)^{2}
             +\dfrac{1}{y}\dfrac{f'(x/y)}{f(x/y)}
             +\dfrac{y}{x^{2}}\dfrac{f(x/y)f'(y/x)}{\left\lbrack f(y/x)\right\rbrack^{2}}.\nonumber
\end{align}

The scalar curvature is given by the linear combination of the
functions $sh_{i}(x,y)$:
$$r(D)=sh_{1}(x,y)-\frac{1}{2}sh_{2}(x,y)+2sh_{3}(x,y)-sh_{4}(x,y).$$

The result is the following:
\begin{align}
r(D)=&2\dfrac{2yf(x/y)-1}{(x-y)^{2}}+6\dfrac{2xf'(x/y)-yf(x/y)}{y(x-y)f(x/y)}
      -\dfrac{1}{2}\dfrac{x(8+3y)}{y^{3}}\left(\dfrac{f'(x/y)}{f(x/y)}\right)^{2}
      -\dfrac{3}{2}\dfrac{y}{x^{2}}\left(\dfrac{f'(y/x)}{f(y/x)}\right)^{2}\nonumber\\[1em]
     &+\dfrac{(3+x)f'(x/y)}{y^{2}f(x/y)}+2\dfrac{xf''(x/y)}{y^{3}f(x/y)}-\dfrac{f'(y/x)}{xf(y/x)}
      -\dfrac{2f'(x/y)f'(y/x)}{x^{2}\left\lbrack f(y/x)\right\rbrack^{2}}+\dfrac{3}{2}.\label{eq:r(f)}
\end{align}

The eigenvalues of $D$ can be expressed by $a$ as
$$\la_{1}=\frac{1+a}{2}\qquad \la_{2}=\frac{1-a}{2}.$$

Substituting these into the previous formula and collecting the
terms we get Eq. (\ref{eq:r(a)}).
\end{proof}

Since the scalar curvature formula Eq. (\ref{eq:r(a)}) is a rather
complicated one it is worth mentioning that there is a completely
different proof (which is based on the subsection 2.1) for Theorem
2.3.

\begin{proof}[Proof 2]
There is another parametrization of the state as it was mentioned
in the introduction.
Let us use the following parametrization for
the $2\times 2$ density matrices:
$$\frac{1}{2}\left(\begin{array}{cc}
1+r\cos\theta & (r\sin\theta\cos\phi)+i(r\sin\theta\sin\phi) \\
(r\sin\theta\cos\phi)-i(r\sin\theta\sin\phi) & 1-r\cos\theta
\end{array}\right),$$
where $(r,\theta,\phi)$ denote the spherical coordinates, but now
$0\leq r<1$. In this case the metric is:
$$ds^{2}=\frac{1}{1-r^{2}}dr^{2}+
  \frac{r^{2}}{(1+r)f\left(\frac{1-r}{1+r}\right)}d\theta^{2}+
  \frac{r^{2}\sin^{2}\theta}{(1+r)f\left(\frac{1-r}{1+r}\right)}d\phi^{2}.$$
Let us use the order $(r,\theta,\phi)$ for the coordinates. (For
example $\partial_{2}t(r,\theta,\phi)$ denotes the partial
derivative of $t(r,\theta,\phi)$ with respect to $\theta$.) The
$g_{ik}$ metric can be written in the form
$g_{ik}=\delta_{ik}\alpha_{i}(r,\theta,\phi)$. The identities
$$\partial_{2}\alpha_{1}=\partial_{3}\alpha_{1}=0,\quad
  \partial_{2}\alpha_{2}=\partial_{3}\alpha_{2}=0,\quad
  \partial_{3}\alpha_{3}=0$$
will simplify the computation. The Christoffel symbols of the
second kind for this Riemannian manifold is
\begin{equation}
\Gamma_{ij}^{..m}=\sum_{k=1}^{3}
  \frac{1}{2}g^{km}(\partial_{i}g_{jk}+\partial_{j}g_{ik}-\partial_{k}g_{ij}),\label{eq:Christoffel}
\end{equation}
where $g^{ij}$ denotes the inverse matrix of $g_{ij}$. Since
$\Gamma_{ij}^{..m}=\Gamma_{ji}^{..m}$, there are only seven
nonzero independent Christoffel symbols in this case
\begin{align}
 &\Gamma_{1,1}^{..1}=\dfrac{r}{1-r^{2}},\qquad
  \Gamma_{2,2}^{..1}=\dfrac{-r(1-r)}{2(1+r)^{2}f(c(r))}
                     \left(r^2+3r+2+2r\dfrac{f'(c(r))}{f(c(r))}\right),\qquad
  \Gamma_{3,3}^{..1}=\sin^{2}\theta\ \Gamma_{2,2}^{..1},\nonumber\\[1em]
&\Gamma_{1,2}^{..2}=\dfrac{-f(c(r))}{r^{2}(1-r)}\Gamma_{2,2}^{..1},\qquad
  \Gamma_{3,3}^{..2}=-\sin\theta \cos\theta,\qquad
  \Gamma_{1,3}^{..3}=\Gamma_{1,2}^{..2},\qquad
  \Gamma_{2,3}^{..3}=\dfrac{\cos\theta}{\sin\theta},\label{eq:Christoffel is}
\end{align}
where $c(r)=\frac{1-r}{1+r}$.

The Riemannian curvature tensor is given by the equation
\begin{equation}
R_{ijkl}=\sum_{n=1}^{3}g_{ln}
  \left(\partial_{i}\Gamma_{jk}^{..n}-\partial_{j}\Gamma_{ik}^{..n}+
  \sum_{m=1}^{3}\bigl(\Gamma_{jk}^{..m}\Gamma_{im}^{..n}-\Gamma_{ik}^{..m}\Gamma_{jm}^{..n} \bigr)
 \right).\label{eq:Riemann}
\end{equation}
Since $R_{ijkl}=-R_{jikl}$, $R_{ijkl}=-R_{ijlk}$ and
$R_{ijkl}=R_{klij}$, there are only three nonzero independent
element of the curvature tensor:
\begin{align}
R_{1212}&\displaystyle=\frac{-r}{(1+r)^4(1-r^2)f(c(r))}\left(
   2r(1-r)\frac{f''(c(r))}{f(c(r))}-3r(1-r)\left(\frac{f'(c(r))}{f(c(r))}\right)^{2}\right.\label{eq:Riemann is}\\[1em]
   &\displaystyle\left.+(1+r)(3r-2)\frac{f'(c(r))}{f(c(r))}+\frac{(r^{2}+r+4)(1+r)^{2}}{4}\right)\nonumber\\[1em]
R_{1313}&\displaystyle=\sin^{2}\theta \ R_{1212}\nonumber\\[1em]
 R_{2323}&\displaystyle=\frac{r^{2}(1-r)\sin^{2}\theta}{(1+r)^4\left\lbrack f(c(r))\right\rbrack^2}\biggl(
   r(r+2)\frac{f'(c(r))}{f(c(r))}+\frac{r^{2}}{1+r}\left(\frac{f'(c(r))}{f(c(r))}\right)^{2}
   -\frac{(1+r)^3}{1-r}f(c(r))\nonumber\\[1em]
   &\displaystyle+\frac{(1+r)(2+r)^{2}}{4}\biggr).\nonumber
\end{align}

The Ricci curvature tensor is
\begin{equation}
\Ric_{ij}=\sum_{k,l=1}^{3}g^{kl}R_{lijk}.\label{eq:Ricci}
\end{equation}
It is symmetric $\Ric_{ij}=\Ric_{ji}$, and it has three nonzero
elements:
\begin{align}
\Ric_{1,1}&\displaystyle =\frac{1}{(1+r)^4}\left(
   4\frac{f''(c(r))}{f(c(r))}-6\left(\frac{f'(c(r))}{f(c(r))}\right)^{2}
   +\frac{2(1+r)(3r-2)}{(1-r)} \frac{f'(c(r))}{f(c(r))}\right.\label{eq:Ricci is}\\[1em]
   &\displaystyle\left.+\frac{(r^{2}+r+4)(1+r)^{2}}{2r(1-r)}\right)\nonumber\\[1em]
\Ric_{2,2}&\displaystyle =\frac{r^{2}(1-r)}{(1+r)^4f(c(r))}\left(
   2\frac{f''(c(r))}{f(c(r))}-4\left(\frac{f'(c(r))}{f(c(r))}\right)^{2}
   +\frac{(1+r)(r^{2}+4r-4)}{r(1-r)} \frac{f'(c(r))}{f(c(r))}\right.\nonumber\\[1em]
   &\displaystyle\left.+\frac{(1+r)^4}{r^{2}(1-r)}f(c(r))
   +\frac{(r^{3}+2r^{2}+2r-2)(1+r)^{2}}{2r^{2}(1-r)}\right)\nonumber\\[1em]
\Ric_{3,3}&\displaystyle=\sin^{2}\theta\ \Ric_{2,2}.\nonumber
\end{align}

The scalar curvature at point $D$ is
\begin{equation}
r(D)=\sum_{i,j=1}^{3}g^{ij}\Ric_{ji}.\label{eq:Scal from Ricci}
\end{equation}
Computing $r(D)$ we get Eq. (\ref{eq:r(a)}).
\end{proof}

The state $D$ is maximally mixed if its eigenvalues are equal, in this case $a=0$.
So the scalar curvature has local minimum or maximum at the maximally mixed
state if and only if the function $r(a)$ has local minimum or maximum at
the origin.

\subsection{Curvature formula at the origin and Radon measures}

To find an operator monotone function $f$ such that the scalar curvature
has local minimum at the origin we start from the following representation
theorem in Ref. 19.

The map $\mu\mapsto f$ defined by
$$f(x)=\int_{0}^{\infty} \frac{x(1+t)}{x+t}\ d\mu(t)
  \qquad \mbox{for}\quad x>0$$
establishes an affine isomorphism from the class of positive Radon measures
$\lbrack 0,\infty\rbrack$ onto the class of operator monotone functions.

We use a modified version of the previous theorem Ref. 20.

\begin{theorem} The map $\mu\mapsto f$, defined by
\begin{equation}
f(x)=\int_{0}^{1}\frac{x}{(1-t)x+t}\ d\mu(t),\qquad\mbox{for}\quad
x>0,\label{eq:f repr}
\end{equation}
establishes a bijection between the class of positive Radon measures on
$\lbrack 0,1 \rbrack$ and the class of operator monotone functions.
\end{theorem}

From this representation we get that
$$xf(x^{-1})=\int_{0}^{1}\frac{x}{(1-t)+tx}\ d\mu(t)
            =\int_{0}^{1}\frac{x}{(1-t)x+t}\ d\mu(1-t)\ .$$
Thus $f(x)=xf(x^{-1})$ holds iff $\mu(\lbrack 0,t\rbrack
)=\mu(\lbrack 1-t,1\rbrack)$ for all $t\in\lbrack 0,1\rbrack$ and
the $f(1)=1$ normalization means that $\mu(\lbrack 0,1\rbrack)=1$.
Let $T$ denote the set of all positive Radon measures on the
$\lbrack 0,1\rbrack$ interval such that $\mu(X)=\mu(1-X)$ for
every measurable $X$ subset of $\lbrack 0,1\rbrack$ and
$\mu(\lbrack 0,1\rbrack)=1$. Theorem 2.1 and 2.4 imply that there
is bijective correspondence between monotone metrics and $T$.

\section{SCALAR CURVATURE}

\subsection{Scalar curvatures with local minimum at the origin}

For detailed verification of Theorem 3.1 and 3.3 the \textit{Maple} program was used.
The Maple worksheet, containing these proofs is available at Ref. 21.

\begin{theorem}
The series expansion of $r(a)$ at the origin leads to the
\begin{eqnarray}
        &r(a) =  (6+36f''(1))
         +a^{2}\left(\frac{100}{3}f^{(4)}(1)-140f''(1)
                     -120f''(1)^{2} \right)
         +a^{4}\Bigl(352f''(1)^{3}  \label{eq:r(a) series}\\
        &+616f''(1)^{2}+1092f''(1)-\frac{1288}{3}f^{(4)}(1)
                     +\frac{392}{45}f^{(6)}(1)-160f''(1)f^{(4)}(1) \Bigr)
         +O(a^{6})\nonumber
\end{eqnarray}
approximation.
\end{theorem}

\begin{proof}

From Eq. (\ref{eq:r(a)}) one may expect that the $1/a$ and
$1/a^{2}$ type divergences occur in this expansion but the
behavior derivatives of $f$ not allow this. It is obvious that
$r(a)=r(-a)$ from symmetric reasons (not from the formula!) this
means that the coefficient of $a^{(2n+1)}$ will be zero for all
$n\in\mathbb{N}$. We proof this series expansion only up to the
order $O(a^4)$ because the coefficient of $a^{4}$ can be derived
in a similar way, but it needs more complicated formulas. Through
the computation we will use the identities
$$f'(1)=\frac{1}{2},\quad f^{(3)}(1)=-\frac{3}{2}f^{(2)}(1),\quad
  f^{(5)}(1)=-\frac{15f^{(4)}(1)+60f^{(3)}(1)+60f^{(2)}(1)}{2}$$
which come from the equations $f(x)=xf(1/x)$ and $f(1)=1$. We
consider the scalar curvature as a sum of five functions according
to the Eq. (\ref{eq:r(a)}). The series expansion of the summands
can be computed in elementary way, but the intermediate formulas
are rather complicated. The series expansions of the five summands
from Eq. (\ref{eq:r(a)}) after simplifications are the following.
$$\begin{array}{rl}
&\mbox{1st:}\ -\frac{7}{2}+7\left(4f''(1)+1\right)\cdot a
            -7\left(8f''(1)^{2}+4f''(1)+1 \right)\cdot a^2 \\[1em]
&\mbox{2nd:}\ -6\cdot\frac{1}{a}+\left(24f''(1)+13\right)-\left(28f''(1)+12\right)\cdot a
             +\left(16f^{(4)}(1)-48f''(1)^{2}-52f''(1)+12 \right)\cdot a^{2} \\[1em]
&\mbox{3rd:}\ 8f''(1)+\left(16f^{(4)}(1)-16f''(1)^{2}-56f''(1) \right)\cdot a^{2} \\[1em]
&\mbox{4th:}\ 2\cdot\frac{1}{a^{2}}+4f''(1)
             +\left(\frac{4}{3}f^{(4)}(1)-4f''(1) \right)\cdot a^{2}\\[1em]
&\mbox{5th:}\ -2\cdot\frac{1}{a^{2}}+6\cdot\frac{1}{a}-5+5\cdot a-5\cdot a^{2} \\[1em]
\end{array}$$
The sum of these expansions leads to Eq. (5) in this theorem.

\end{proof}

Combining Eqs. (\ref{eq:f repr}) and (\ref{eq:r(a) series}) we
conclude that the scalar curvature has local minimum at the origin
if
\begin{equation}
12\left(\int_{0}^{1}t(1-t)\ d\mu(t)\right)^{2}-
\int_{0}^{1}t(t-1)(20t^{2}-40t+13)\ d\mu(t)<0
\label{eq:min.condition}
\end{equation}
holds for a $\mu\in T$. The scalar curvature at the origin is given by
$$6+72\int_{0}^{1} (t^{2}-t)\ d\mu(t).\label{eq:r at origin}$$
It has maximum when $\mu=(1/2)\delta_{0}+(1/2)\delta_{1}$, the
corresponding operator monotone function is $f(x)=\frac{1+x}{2}$
and then $r(0)=6$. It has minimum when $\mu=\delta_{1/2}$, the
corresponding operator monotone function is $f(x)=\frac{2x}{1+x}$
and then $r(0)=-12$.

The measure $\mu\in T$ can be transformed into a probability
measure $\mu'$ on the $\lbrack 0,1 \rbrack$ interval such that:
$$\int_{0}^{\frac{1}{2}}t(1-t)\ d\mu(t)=\frac{1}{8}\int_{0}^{1}x\ d\mu'(x)$$
because the $4t(1-t)$ function maps the $2\mu\vert_{\lbrack 0,1/2\rbrack}$
measure into a probability measure on $\lbrack 0,1\rbrack$.
If $\lambda$ denotes the Lebesgue-measure and
$$\mu(t)\vert_{\left\lbrack 0,\frac{1}{2}\right\rbrack}
=\rho(t)\ d\lambda(t)+\sum a_{i}\delta_{p_{i}}$$
then
$$\mu'(x)=\frac{1}{2}\rho\left(\frac{1-\sqrt{1-x}}{x}\right)\frac{1}{\sqrt{1-x}}\ d\lambda(x)
  +\sum 2a_{i}\delta_{4p_{i}(1-p_{i})}.\label{eq:mu acute}$$
There is one to one correspondence between probability measures on
$\lbrack 0,1 \rbrack$ and $T$. Let $m_{\mu}$ denote the
expectation $\sigma^{2}_{\mu}$ the variance and $E_{n,\mu}$ the
n-th momentum of the $\mu'$ measure. Using Eq. (\ref{eq:f repr})
and the previous notation one can check the following equalities
$$f''(1)=-\frac{m_{\mu}}{2} \quad
  f^{(4)}(1)=-3m_{\mu}+\frac{3}{2}E_{2,\mu}\quad
  f^{(6)}(1)=-90m_{\mu}-\frac{45}{4}E_{3,\mu}+90E_{2,\mu}$$
Substituting this into the approximation
  Eq. (\ref{eq:r(a) series})
 one get the following theorem.

\begin{theorem} If for a measure $\mu\in T$ the inequality
 $$m_{\mu}(3-2m_{\mu})<5\sigma_{\mu}^{2}$$
holds or if
$$m_{\mu}(3-2m_{\mu})=5\sigma_{\mu}^{2}\quad \mbox{and}\quad
  -44m_{\mu}^{3}+70m_{\mu}^{2}+114m_{\mu}<98E_{3,\mu}$$
then the scalar curvature of the metric induced by the measure
$\mu$ by the Eq. (\ref{eq:f repr}) has local minimum at the
origin.
\end{theorem}

We give examples for monotone metrics which satisfies the previous
conditions so the scalar curvature of them has local minimum at
the maximally mixed state.

\begin{theorem} Let
$$\frac{7-\sqrt{7}}{14}<p\leq\frac{1}{2}$$
and
\begin{equation}
h(p)=
\frac{\sqrt{14p^{2}-14p+4+\sqrt{-640p^{4}+1280p^{3}-880p^{2}+240p+9}}}{2\sqrt{7}}
 \label{eq:h(p)}
\end{equation}
and $0\leq q<\frac{1}{2}-h(p)$.
Then the scalar curvature of the $\MM$ manifold coming from the
operator monotone function
\begin{equation}
f(x)=\frac{x}{4}\left( \frac{1}{px+1-p}+\frac{1}{(1-p)x+p}
     +\frac{1}{qx+1-q}+\frac{1}{(1-q)x+q} \right)
\label{eq:f(x,q)}
\end{equation}
has local minimum at the origin.
\end{theorem}

\begin{proof}

First one can try to find a $\mu\in T$ measure such that Eq.
(\ref{eq:min.condition}) holds in
\begin{equation}
\mu_{p}=\frac{1}{2}\delta_{p}+\frac{1}{2}\delta_{1-p}
  \label{eq:mu 1 Dirac}
\end{equation}
form where $\delta_{p}$ is a Dirac-measure. Let
\begin{equation}
t_{\mu}:=12\left(\int_{0}^{1}t(1-t)\ d\mu(t)\right)^{2}-
  \int_{0}^{1}t(t-1)(20t^{2}-40t+13)\ d\mu(t)
  \label{eq:t(mu) 1 Dirac}
\end{equation}
and $t(p)=t_{\mu_{p}}$. We get that
$$t(p)=p(1-p)(8p^{2}-8p+3).$$
For all $p\in\lbrack 0,1/2 \rbrack$ we have $t(p)>0$.
This means that the scalar curvature has local maximum at the origin for all
$\mu_{p}$ measures.

Let $p\in\lbrack 0,1/2 \rbrack$, $q\in\lbrack p,1/2 \rbrack$ and
\begin{equation}
\mu_{p,q}=\frac{1}{4}\delta_{p}+\frac{1}{4}\delta_{q}+
            \frac{1}{4}\delta_{1-p}+\frac{1}{4}\delta_{1-q}.\label{eq:mu 2 Dirac}
\end{equation}
Let $t(p,q)=t_{\mu_{p,q}}$ then
\begin{equation}
t(p,q)=-7(p^{4}+q^{4})+14(p^{3}+q^{3})-6pq(p+q-pq-1)-\frac{17}{2}(p^{2}+q^{2})
 +\frac{3}{2}(p+q).\label{eq:t(mu) 2 Dirac}
\end{equation}
After substituting into Eq. (\ref{eq:t(mu) 2 Dirac}) the
\begin{equation}
p=\frac{v+\sqrt{v^{2}-4u}}{2}\qquad q=\frac{v-\sqrt{v^{2}-4u}}{2}
  \label{eq:p(u,v) q(u,v)}
\end{equation}
formulas one derives that
\begin{equation}
t(u,v)=-8u^{2}+(28v^{2}-48v+23)u-
\left(7v^{4}-14v^{3}+\frac{17}{2}v^{2}-\frac{3}{2}v\right).\label{eq:t(u,v)}
\end{equation}
The equation $t(u,v)=0$ has two solutions for a given $v$. Taking
into account that $u=pq$ we get the condition $0<u<\frac{1}{4}$.
The only solution of the equation $t(u,v)=0$ which fulfills this
condition is
\begin{equation}
u(v)=\frac{7}{4}v^{2}-3v+\frac{23}{16}
  -\frac{1}{16}\sqrt{560v^{4}-2240v^{3}+3320v^{2}-2160v+529}.\label{eq:u(v)}
\end{equation}
If the parameter $p$ is given then $q$ can be computed from the
equation $u(p+q) = pq$. There are four solutions for $q$ but only
one of them is admissible
\begin{equation}
q(p)=\frac{1}{2}-\frac{1}{14}\sqrt{84p^{2}-84p+28+7\sqrt{-640p^{4}+1280p^{3}-880p^{2}+240p+9}}
  \label{eq:q(p)}
\end{equation}
because of the conditions for $q$. This equation gives positive
parameter $q$ if
$$ \frac{7-\sqrt{7}}{14}<p\leq\frac{1}{2}.$$
One can check that if $0<q<q(p)$ then the function $t(p,q)$ is
negative. Then we use Eq. (\ref{eq:f repr}) defining a desired
$f(z)$ operator monotone function from the $\mu_{p,q}$ measures.

\end{proof}

If we choose $\frac{7-\sqrt{7}}{14}<p\leq\frac{1}{2}$ arbitrary
and $q=0$ in the previous theorem then we get, that the scalar
curvature coming from the operator monotone function
$$f(x)=\frac{x}{4}\left(\frac{1}{(1-p)x+p}+\frac{1}{px+1-p}+\frac{1}{x}+1 \right)$$
has local minimum at the origin. In this case
series expansion of the scalar curvature at the origin is
$$r(a)=\left(\frac{9}{2}-36p(1-p)\right)-20p(1-p)(14p^{2}-14p+3)\cdot a^{2}+O(a^{4}).$$
One can prove that the minimum at the origin is not only local but global for these
functions. The greatest value of the scalar curvature in this case is
$$r(1)=\frac{7}{2}+\frac{1}{p(1-p)}.$$

Here some other examples for operator monotone functions, such that the scalar
curvature derived from them has local, but not global minimum at the maximally
mixed state.
$$f(x)=\frac{x}{4} \left(\frac{4}{x+1}+\frac{50}{x+49}+\frac{50}{49x+1}\right),
\quad
f(x)=\frac{250x}{999x+1}+\frac{250x}{x+999}+\frac{x}{x+1}.$$
Numerical computations suggest that the scalar curvature of the
Riemannian metric of a three level quantum system induced by the second
function has local minimum at the maximally mixed state.

\section{CONCLUSIONS}

The Riemannian metrics so far studied on the manifold $\MN$ come from
special operator monotone functions according to Theorem 2.1.
The metric carries all differential geometrical properties of the manifold, this means
that from a suitable function $f$ one can derive all geometrical quantities of the manifold.
One of the basic phenomenological problem is to give physical
interpretation of differential geometrical quantities.

One can expect that the greatest statistical uncertainty should belong
to the most mixed states. This expectation means that the scalar curvature
of a Riemannian metrics should have a global
maximum at the maximally mixed state. We gave several examples for suitable operator monotone
functions such that the derived scalar curvatures do not fulfill this expectations
and have even a local minimum at the maximally mixed state.

The Kubo-Mori (or Bogoliubov) metric comes from the function $f(x)=\frac{x-1}{\log x}$.
This is one of the statistically most relevant metrics. It was
conjectured in Ref. 17 that the scalar curvature of this metric is
monotone in the following sense. If $D_{1},D_{2}\in\MN$
and $D_{1}$ is more mixed than $D_{2}$ then $r(D_{1})\geq r(D_{2})$.

\bigskip

{\bf Acknowledgement.} I would like to thank Dr. D\'enes Petz for
stimulating discussions and valuable remarks.
  This work partially supported by OTKA T32374 and OTKA T43242.

\begin{enumerate}

\item[$^{\lbrack 1\rbrack}$] C. R. Rao,
``Information and accuracy attainable in the estimation of statistical parameters,''
Bulletin of the Calcutta Mathematical Society {\bf 37}, 81--91 (1945).

\item[$^{\lbrack 2\rbrack}$] S. Amari,
\textit{Differential-geometrical methods in statistics},
Lecture Notes in Statistics, Springer, Berlin-New York, 1985.

\item[$^{\lbrack 3\rbrack}$] R. F. Streater, ``Classical and
quantum info-manifolds,'' S\'urikaisekikenky\'usho K\'oky\'uroku
{\bf 1196}, 32--51 (2001), math-ph/0002050.

\item[$^{\lbrack 4\rbrack}$] R. A. Fisher,
``On the mathematical foundations of theoretical statistics,''
Phil. Trans.,  A, {\bf 222}, 309--368 (1922).

\item[$^{\lbrack 5\rbrack}$] N. N. Cencov,
\textit{Statistical decision rules and optimal inference},
Translations of Mathematical Monograph, 53, 1982.

\item[$^{\lbrack 6\rbrack}$] E. A. Morozova and N. N. Chentsov,
``Markov invariant geometry on state manifolds,'' (Russian)
Translated in J. Soviet Math. {\bf 56}, 2648--2669 (1991).
Current problems in mathematics. Newest results,
Itogi Nauk Tekhn {\bf 36}, 69--102 (1990) (in Russian).

\item[$^{\lbrack 7\rbrack}$] D. Petz and Cs. Sud\'ar,
``On the curvature of a certain Riemannian space of matrices,''
J. Math. Phys. {\bf 37}, 2662--2673 (1996).

\item[$^{\lbrack 8\rbrack}$] D. Petz
``Covariance and Fisher information in quantum mechanics,''
J. Phys. A {\bf 35}, 929--939 (2002).

\item[$^{\lbrack 9\rbrack}$] J. Dittmann
``The scalar curvature of the Bures metric on the space of density matrices,''
Journal of Geometry and Physics {\bf 31}, 16--24 (1999).

\item[$^{\lbrack 10\rbrack}$] J. Dittmann,
``On the curvature of monotone metrics and a conjecture concerning the Kubo-Mori metric,''
Linear Algebra Appl., {\bf 315}, 83--112 (2000).

\item[$^{\lbrack 11\rbrack}$] J. Dittmann,
``On the Riemannian geometry of finite dimensional mixed states,''
Sem. S. Lie, {\bf 3}, 73--87 (1993).

\item[$^{\lbrack 12\rbrack}$] P. W. Michor, D. Petz and A. Andai,
``On the curvature of a certain Riemannian space of matrices,''
Infinite Dimensional Analysis, Quantum Probability, {\bf 3},
199--212 (2000).

\item[$^{\lbrack 13\rbrack}$] P. M. Alberti, A. Uhlmann,
``Stochasticity and partial order. Doubly stochastic maps and
unitary mixing,'' in {\textit Mathematics and its Applications, 9}
D. Reidel Publishing Co., Dordrecht-Boston, Mass., 1982. 123pp.

\item[$^{\lbrack 14\rbrack}$] R. Bhatia,
\textit{Matrix Analysis},
Springer-Verlag, 1997.

\item[$^{\lbrack 15\rbrack}$] F. Hiai, D. Petz and G. Toth,
``Curvature in the geometry of canonical correlation,''
Studia Scientiarum Mathematicarum Hungarica {\bf 32}, 235--249 (1996).

\item[$^{\lbrack 16\rbrack}$] S. Gallot, D. Hulin and J. Lafontaine,
\textit{Riemannian Geometry}, Springer-Verlag, 1987,1990.

\item[$^{\lbrack 17\rbrack}$] D. Petz,
``Geometry of canonical correlation on the state space of a quantum system,''
J. Math. Phys. {\bf 35}, 780--795 (1994).

\item[$^{\lbrack 18\rbrack}$] D. Petz,
``Monotone metrics on matrix spaces,''
Linear Algebra Appl., {\bf 244}, 81--96 (1996).

\item[$^{\lbrack 19\rbrack}$] F. Kubo and T. Ando,
``Means of positive linear operators,''
Math. Ann. {\bf 246}, 205--224 (1980).

\item[$^{\lbrack 20\rbrack}$] P. Gibilisco and T. Isola,
``Monotone metrics on statistical manifolds of density matrices by geometry of
non-commutative $L^{2}$-Spaces,'' in
{\textit Disordered and Complex Systems}
(AIP Conf. Proc., 553, Amer. Inst. Phys., Melville, NY, 2001), 129--140.

\item[$^{\lbrack 21\rbrack}$] A. Andai,
\textit{http://www.math.bme.hu/\~{}andaia/maple/}.

\end{enumerate}

\end{document}